# Formalism of Requirements for Safety-Critical Software: Where Does the Benefit Come From?

Ibrahim Habli and Andrew Rae
Department of Computer Science
University of York
York, United Kingdom
Ibrahim.Habli@york.ac.uk
Andrew.Rae@york.ac.uk

*Abstract*— Safety and assurance standards often rely on the principle that requirements errors can be minimised by expressing the requirements more formally. Although numerous case studies have shown that the act of formalising previously informal requirements finds requirements errors, this principle is really just a hypothesis. An industrially persuasive causal relationship between formalisation and better requirements has yet to be established. We describe multiple competing explanations for this hypothesis, in terms of the levels of precision, re-formulation, expertise, effort and automation that are typically associated with formalising requirements. We then propose an experiment to distinguish between these explanations, without necessarily excluding the possibility that none of them are correct.

## I. INTRODUCTION

There are two main strategies available for evaluating the effectiveness of a software standard. Either the standard as a whole can be compared to alternatives, or the individual practices within the standard can be separately evaluated. This position paper describes how the second approach could be applied, using one specific practice as an illustrative example.

Standards for the development of safety-critical software typically contain rules for how the software requirements should be represented (e.g. [1] [2]). The underlying principle is that requirements errors can be minimised by expressing the requirements more formally [3]. This principle is really just a hypothesis. Whilst there are numerous case studies showing that the act of formalising previously informal requirements finds requirements errors [4] [5], an industrially persuasive causal relationship between formalisation and better requirements has yet to be established.

Understanding this causal relationship is important for a number of reasons:

1. From a practical point of view, formalisation has significant cost and effort implications. Using formal requirements is a significant intellectual investment

2. Formalisation addresses a substantial ideological divide, with different communities and stakeholders taking different entrenched positions

3. There has been a lot of research effort, but not enough studies that directly address the "should we do it" question – this question is mostly secondary, and therefore addressed as a case-study side-effect rather than by decisive experiment [9].

## II. THE QUESTION TO BE ADDRESSED

We here have an empirical observation, that application of formalisation methods to requirements finds and removes requirements errors. We have multiple competing hypotheses to explain this observation.

The first explanation is that the errors are found as a result of the degree of precision provided by the increased formality in representation. For example, four different levels of formalism can be considered: free text, structured text [6], semi-formal models [7] and mathematically-based notations [4].

The second explanation is that the errors are found as a result of re-formulation of the requirements, regardless of the nature of the new form. This explanation is suggested by the fact that all requirements formalisation methods seem capable of finding requirements errors.

The third explanation is that the errors are found as a result of expertise. Formal methods are applied by highly qualified practitioners expert in those methods, and presumably expert in spotting requirements errors.

The fourth explanation is that the error removal is a simple result of effort expended, independent of the method or the expertise.

The fifth explanation is that the errors are found as a result of the increase of automation (e.g. the use of modelling tools for model-based languages or theorem-provers or model-checking for mathematically-based languages [8]).

Our challenge is to design an experiment to distinguish between these explanations, without excluding the possibility that none of them are correct.

## III. THE EXPERIMENT

Our experimental method is based around the principle of dose-response as an indicator of causality. Given a particular set of requirements, participants will be set a task of reformulating the requirements and identifying problems. Performance will be determined according to the number of distinct problems identified with the requirements. Five parameters of the task will be set corresponding to the five explanations.

Since it is possible that more than one of the explanations is a causal factor, and that the explanations may in fact influence each other, an ideal experiment should consider permutations of the parameters rather than simply hold three steady and adjust the last parameter.

Briefly, the parameters are:

1. Degree of formalism of the new form of the requirements;
2. Difference between the old and new form of requirements;
3. Expertise of the participants;
4. Time available for the task; and
5. Degree of automation available for the new form of the requirements.

An explanation will be determined to have causal power if increasing the parameter associated with that explanation improves performance on the task independently of the other parameters.

## IV. THREATS TO VALIDITY

The internal validity of this experimental approach relies on the causal relationships to be, if not linear, at least monotonic. If increasing any of the parameters by a small amount improves the task performance, but increasing it by a large amount has negligible or negative effect, this will be hard to detect.

The most obvious challenges to the external validity of this experiment will relate to the particular set of requirements presented as a challenge problem. If finding the errors is too easy or too hard, this may decrease or increase the utility of formalisation. Formalisation may be good at finding subtle errors, but no more useful than informal re-formulation for more obvious errors. More broadly, there may be particular types of systems or requirements for which formality is more useful. Formality may be very effective for control systems, but not particularly useful for simple input-output systems or complex adaptive systems, for example.

The question may be successfully answered in an experimental context, but the results may not be industrially applicable. Formalisation requires a precise description of the system. What about if the requirements are inherently fluid? Freezing the requirements may be harmful. There is a trade-off between requirements consistency and requirements completeness. From a safety point of view we care about "global correctness" (e.g. fitness against intent rather than specification), which includes both consistency and completeness and covers both requirements verification and validation.

This threat is inherent to our "individual practice" approach. If the benefit or detriment of a standard is an emergent property of a suite of practices, or even simply of *having* a defined suite of practices, then evaluating the efficacy of individual practices will not appropriately evaluate the standard, even if all of the practices are rigorously addressed in isolation. On the other hand, if benefit or detriment is inherent to particular practices, then only an individual practice approach will give a true understanding of the efficacy of standards.

## V. PRACTICALITY

What is a reasonable sized task to test the usefulness of requirements reformulation? What is a reasonable amount of time for the task? How large does each group of participants need to be to ensure the results are representative? How many different combinations of parameters do we need to test? At this point, we do not know the answers to these questions.

Our suggested approach is peer-design of the experiment. This requires finding a "champion" for each of the candidate explanations, who has previously expressed support for that explanation, and currently stands by that view. If each of the champions agrees that the task set and the means of measuring performance is a fair way of comparing the explanations, this increases confidence in the usefulness of the experiment.

One pragmatic approach would be to gradually increase the expertise of the participants, using students as initial subjects. This would allow the practicality of the approach to be explored at the same time as refining the parameter combinations to be usefully tested with higher expertise groups.